\documentclass[preprint,nofootinbib]{revtex4}

\usepackage{ulem}
\usepackage{amssymb}
\usepackage{amsmath}
\usepackage{amsfonts}
\usepackage{lscape}
\usepackage{epsfig}
\usepackage{color}
\def\beq{\begin{equation} }
\def\eeq{\end{equation} }
\def\ba{\begin{eqnarray} }
\def\ea{\end{eqnarray} }
\def\nn{\nonumber }

\begin{document}
%\draft
\title{Fuzzy de Sitter space from  kappa-Minkowski space in matrix basis}
\author{Danijel Jurman}
\email{djurman@irb.hr}
\affiliation {Division of Theoretical Physics, Rudjer Bo\v skovi\'c Institute,
Bijeni\v cka 54, 10000 Zagreb, Croatia}

\vspace{2cm}

\begin{abstract}
We consider the Lie group $\mathbb{R}^D_\kappa$ generated by the Lie algebra of $\kappa$-Minkowski space.
Imposing the invariance of the metric under the pull-back of diffeomorphisms induced by right translations in the group, we show that a
unique right invariant metric is associated with $\mathbb{R}^D_\kappa$. This metric coincides with the metric of de Sitter space-time.
We analyze the structure of unitary representations of the group $\mathbb{R}^D_\kappa$ relevant for
the realization of the non-commutative $\kappa$-Minkowski space by embedding into $(2D-1)$-dimensional Heisenberg algebra. 
Using a suitable set of generalized coherent states, we select the particular Hilbert space and realize
the non-commutative $\kappa$-Minkowski space as an algebra of the Hilbert-Schmidt operators.
We define dequantization map and fuzzy variant of the Laplace-Beltrami operator such that dequantization map relates 
fuzzy eigenvectors with the eigenfunctions of the Laplace-Beltrami operator on the half of de Sitter space-time.
\end{abstract}

%\pacs{}
\maketitle

\section{introduction}
Recent analysis \cite{Kim:2011ts} of the matrix model that has been proposed as a nonperturbative 
formulation of type IIB superstring theory \cite{Ishibashi:1996xs}
showed that  3+1-dimensional non-commutative $\kappa$-Minkowski  
space is compatible with the solution discovered by a numerical analysis in \cite{Kim:2011cr} and interpreted as 
an expanding universe. In general, such non-commutative spaces represent a realization of an old idea, proposed by W. Heisenberg 
and realized by H. S. Snyder \cite{Snyder:1946qz}, that
space-time has a structure which in the currently available experiments manifests itself as a smooth manifold,
whereas its fundamental description needs some modified notion of space-time.

In the particular model \cite{Ishibashi:1996xs}, as well as in  similar dynamical matrix models of Yang-Mills type,
the structure of space-time is described using matrix geometry \cite{Balachandran:2005ew,madbook}
and emergent gravity \cite{Yang:2009pm, Steinacker:2010rh, Steinacker:2007dq, DelgadilloBlando:2008vi}.
While matrix geometry provides a tool to identify the geometry and field content of the model,
the mechanism of emergent gravity gives rise to the couplings between these fundamental degrees of freedom.
Specifically, coupling of the fields to geometry is defined by derivatives following  
from an expansion around a classical solution.
Whenever derivatives span a Lie algebra, the matrix algebra can be viewed as a quantization of 
the algebra of functions on a certain homogeneous space.
Then, in principle, a quantization map  can be defined  explicitly \cite{Grosse:1993uq}, which might be particularly
beneficial in an attempt to formulate quantized/non-commutative counterparts of models used in  
cosmology or field theory on a curved background \cite{SheikhJabbari:2006bj,Gazeau:2009mi,Stern:2014aqa,Chaney:2015ktw,Buric:2014ika,Buric:2015wta,Chatzistavrakidis:2018vfi,Manolakos:2019fle}.
Therefore, understanding of the quantization map associated to the $\kappa$-Minkowski space 
in terms of matrix geometry is welcome.

Although $\kappa$-Minkowski space is one of the most studied
non-commutative spaces, it has mainly been studied using the algebraic formalism and the structure of Hopf algebra 
(for an overview see \cite{Lukierski:2016vah}).
In this context, several papers \cite{KowalskiGlikman:2003we,KowalskiGlikman:2004tz,Freidel:2006gc} 
pointed out that $\kappa$-Minkowski space is related to de Sitter space of momenta.    
However, the geometry and representation theory of the Lie group generated by the $\kappa$-Minkowski Lie 
algebra have not been studied in detail (as emphasized in \cite{Dabrowski:2010yk}, an exception being Ref. \cite{Agostini:2005mf}).

The purpose of this paper is to fill this gap by considering $\kappa$-Minkowski space 
from the matrix-geometry point of view. A lot of the results needed for such an analysis
are already known in the literature.
We recall these results, interpret them in the context of matrix geometry and apply them 
to unveil the classical geometry associated to non-commutative $\kappa$-Minkowski space. 

\section{Classical geometry of the Lie group generated by the Lie algebra of  $\kappa$-Minkowski space}
Aiming at understanding of the classical geometry associated to  $\kappa$-Minkowski space,
in this section we discuss the geometry of the Lie group generated by the Lie algebra of $\kappa$-Minkowski space.  
More precisely, we consider the group denoted by $\mathbb{R}^{D}_\kappa$ with elements
\ba
g(x^\mu)=g(x^0,x^i)=e^{ix^\mu\hat{x}_\mu},\;x^\mu\in \mathbb{R}, 
\ea
where $\hat{x}_\mu$ are generators of $D$-dimensional $\kappa$-Minkowski space
\ba\label{liealg}
[\hat{x}_0,\hat{x}_k]=-i\hat{x}_k,\;[\hat{x}_j,\hat{x}_k]=0,
\ea
while we use Greek letters for the indices which run from $0$ to $(D-1)$ and Latin letters for the indices which run 
from $1$ to $(D-1)$. 
Group multiplication, given by 
\ba\label{grop}
&&g(a,b^k)g(c,d^k)=g\left(a+c, \frac{\phi(a)b^k+e^{a}\phi(c)d^k}{\phi(a+c)}\right),\\
&&\phi(x)=\frac{e^{x}-1}{x},
\ea
can be derived using the isomorphism of two-dimensional $\kappa$-Minkowski space
with the group of affine transformations of the straight line \cite{vil,gn}.

A glimpse at the upper central (derived) series reveals that $\mathbb{R}^D_\kappa$ is a solvable Lie group.
It has an abelian subgroup $\mathbb{R}^{D-1}$ generated by the $(D-1)$-dimensional ideal of the
 $\kappa$-Minkowski Lie algebra spanned by $\hat{x}_k$'s.
Consequently, the group $\mathbb{R}^D_\kappa$ is the semidirect product 
$\mathbb{R}\ltimes \mathbb{R}^{D-1}$ of two subgroups and each element of the group can be written in the form 
\ba\label{deco}
g=g(t,0)g(0,y^k), \;t,y^k\in\mathbb{R},
\ea 
which can be viewed as a choice of coordinates on the group manifold.
For example, the two aforementioned coordinate patches are related by
\ba\label{transf}
 t=x^0,\;y^k=x^k\phi(-x^0),
 \ea
but with the particular choice (\ref{deco}) the fiber bundle structure of the group manifold is manifest.
Namely, the right translations by elements of subgroup $H$, i.e. multiplication of the Lie 
group elements by elements of the given subgroup $H$ from the right, generate nonintersecting orbits and
induce foliation of the group manifold.  
The particular leaf of the foliation contains all elements which generate the same orbit
reflecting the equivalence relation  $g\sim gh,\;h\in H,\;g\in G$. Leaves are mutually diffeomorphic and,  
as a result, the group manifold can be regarded as a principal bundle with $H$ being the structure group.
The base space is identified as the space $G/H$ of left cosets $gH$, 
while the fiber is identical to the subgroup $H$ as a homogeneous space.
In the case of $\mathbb{R}^{D}_\kappa$, the bundle is trivial.

Having convenient coordinates at hand, we can explicitly express the basis 
of the cotangent space of $\mathbb{R}^D_\kappa$ at each point of the cotangent bundle, that is a frame, in terms of 
right invariant one-forms, i.e. one-forms invariant under the pull-back induced by the right translations 
$L_{g^\prime}:g\mapsto gg^\prime $.
The basis of the space of the right invariant one-forms itself  
and the dual basis of the space of the right invariant vector fields are given by   
\ba\label{lif}
&&\theta^0_R=dt,\;\theta^k_R=e^{t} dy^k,\\
&&e_0^R=\partial_t,\;e_k^R=e^{-t}\partial_k.\nn
\ea
Efficiently, one-forms (\ref{lif}) can be obtained from  the Maurer-Cartan form $\theta_R=(dg)g^{-1}=\theta^\mu_R\hat{x}_\mu$,
a Lie algebra valued right invariant one-form, which satisfies the Maurer-Cartan equation
\ba 
d\theta_{R}=\theta_{R}\wedge \theta_{R},
\ea 
and provides an isomorphism between the right invariant vector fields and the Lie algebra of the group.
Consequently, the right invariant vector fields (\ref{lif}) close the same Lie algebra
commutation relations as the defining generators of $\kappa$-Minkowski space.

Similarly, basis of the space of the left invariant one-forms and dual
basis of the  space of the left invariant vector fields follow from  the left invariant Maurer-Cartan form
\ba\label{rlif} 
\theta^0_L=dt,\;\theta^k_L=dy^k+ y^kdt,\\
e_0^L=\partial_t-y^k\partial_k,\;e_k^L=\partial_k.\nn
\ea 

Since left and right translations commute, taking into account that variation of the vector field $V$ under the diffeomorphisms 
generated by vector field $W$ is given by Lie derivative ${\cal L}_W V=[W,V]$, it is easy to see that vector fields 
which generate right translations are left invariant and vice versa. For arbitrary Lie group, left and right invariant vector fields 
are related by the push-forward of the map $g\to g^{-1}$ which is, for the particular case of the group $\mathbb{R}^{D}_\kappa$ 
and coordinates (\ref{deco}), given by
\ba \label{coorformmap}
e^{it\hat{x}_0} e^{iy^i\hat{x}_i}\to  e^{-iy^i\hat{x}_i} e^{-it\hat{x}_0}=e^{-it\hat{x}_0} e^{-i e^t y^i\hat{x}_i}\Rightarrow t\to -t,\;y^i\to -e^ty^i.
\ea 

Considering $\mathbb{R}^{D}_\kappa$ as a differentiable manifold, it can be supplemented with arbitrary  
metric, but taking into account the group structure it would be natural to demand certain invariance conditions.
Due to the Leibnitz rule for Lie derivative, invariance on the right(left) translations is inbuilt in the  
tensor product of right(left) invariant one-forms and vector fields. For example, one can define right invariant or  
left invariant volume form $\omega_{R}$ and $\omega_L$, respectively  
\ba 
\omega_{R}=e^{(D-1)t} dt\wedge \bigwedge_{k=1}^{D-1}  dy^k,\;\omega_{L}=dt\wedge \bigwedge_{k=1}^{D-1} dy^k,
\ea
which reveals that $\mathbb{R}^{D}_\kappa$ in not unimodular. Generally, for higher rank tensors it is even possible to impose bi-invariance, 
i.e. left invariance and right invariance at the same time. Although bi-invariance of the the metric is natural to demand for semi-simple Lie groups
which posses non-degenerate Killing form, in accordance with Cartan's criterion of solvability, Killing form related to $\mathbb{R}^{D}_\kappa$
is degenerate and cannot be used to define a suitable metric.
Instead, only one type of invariance, right or left, of the metric can be required, which  
on an arbitrary Lie group does not select unique metric \cite{Milnor:1976pu}. 
For example, for any regular symmetric matrix $G$ with elements $g_{\mu\nu}$ there is right invariant metric
\ba\label{rmetric}
ds^2_R=g_{\mu\nu}\theta^\mu_R\theta^\nu_R,
\ea
and, in principle, for different choices of the matrix $G$ metrics (\ref{rmetric}) are not equivalent, but 
in the case of $\mathbb{R}^{D}_\kappa$, any right invariant metric
can be recast  into the metric assigned to the de Sitter space written in planar coordinates
\ba\label{metric}
ds^2_R=dt^2-e^{2t} \delta_{kl}dy^kdy^l.
\ea
Metric (\ref{metric}) can be viewed as the metric induced by the embedding of de Sitter space into $(D+1)$-dimensional Minkowski space
\ba\label{embedding}
&&  X^0=-\sinh{ t} - e^{ t}\delta_{kl}\frac{y^ky^l}{2},\nn\\
&&  X^k= y^k e^{ t},\nn\\
&&  X^D=-\cosh{ t} -e^{t}\delta_{kl} \frac{y^ky^l}{2},
\ea 
where the constraint  $X^0-X^D>0$ implies that coordinates $(t,y^i),\;i\in 1,\cdots, (D-1)$ cover just half of de Sitter space \cite{Kim:2002uz}.
 
Transformation from (\ref{rmetric}) to (\ref{metric}) can be understood if we restrict discussion on the two-dimensional 
$\kappa$-Minkowski space for which any right invariant metric with Lorentzian signature,
up to an irrelevant scale, can be written in the form
\ba\label{2dmet}
&&ds^2_R=\frac{1}{{\tau}^2}\left(\cos{2\theta} d{\tau}^2+2\sin{2\theta} d\tau dx-\cos{2\theta} {dx}^2\right),\\
&&\theta\in <-\frac{\pi}{4},\frac{3\pi}{4}],\;\tau=\exp(-t)\nn.
\ea 
Then, the metric (\ref{2dmet}) can be recast into the desired form (\ref{metric}) by the transformation
\ba
&&\tau^\prime=\frac{1}{\sqrt{|\cos{2\theta}|}}\tau,\;t^\prime=-\ln\tau^\prime\\
&&x^\prime=x\sqrt{|\cos{2\theta}|}-\tau\frac{\sin{2\theta}}{\sqrt{|\cos{2\theta}|}},\;\theta\in<-\frac{\pi}{4},\frac{\pi}{4}>,\nn\\ 
&&x^\prime=x\sqrt{|\cos{2\theta}|}+\tau\frac{\sin{2\theta}}{\sqrt{|\cos{2\theta}|}},\theta\in<\frac{\pi}{4},\frac{3\pi}{4}>.\nn
\ea
In the exceptional cases $\theta=\pi/4,3\pi/4$, the metric is flat and will not be discussed here. 

In order to prove this result in the $D$-dimensional case
one writes the metric as a symmetric matrix with elements $m_{\mu\nu}$.
Choosing suitable $SO(D-1)$ rotation of $y^k$'s,
the "vector" $(m_{01},m_{02},\ldots,m_{0D-1})$ can be put to the form $(\tilde{m}_{01},0,\ldots,0)$.
Then, an analogous transformation as in the two-dimensional case
followed by a $SO(D-1)$ rotation of space-like coordinates 
diagonalizes the metric and appropriate scaling of spacelike coordinates 
gives (\ref{metric}) up to an irrelevant scale. Besides, we note that the Euclidean variant can be treated similarly.

Finally, due to the map (\ref{coorformmap}), left invariant metric
\ba 
ds^2_L=\left(1-\delta_{kl}y^ky^l\right)  dt^2-2\sum_{k=1}^{D-1}dy^k dt-\delta_{kl}dy^kdy^l,
\ea
can be viewed as a pull-back of the right invariant metric (\ref{metric}) with roles of the left and right invariant fields interchanged, 
i.e. left invariant metric is given in terms of a tensor product of left invariant forms, 
while the Killing vectors coincide with the right invariant vector fields.

To summarize,  for $\mathbb{R}^{D}_\kappa$ there is a unique right invariant metric which 
is equivalent to the unique left invariant metric via the pull-back of the map 
$g \to g^{-1}$. With this metric $\mathbb{R}^{D}_\kappa$ is equivalent to the half of de Sitter
space-time. The form  of the metric (\ref{metric}) indicates that  $\mathbb{R}^{D}_\kappa$ can be viewed as 
a homogeneous space with respect to $SO(1,D-1)$ group. 
    
For later purpose, recall that eigenfunctions of the invariant Laplace-Beltrami operator in planar coordinates:
\ba\label{invlap}
\Box_g=\frac{1}{\sqrt{|g|}}\partial_\mu \sqrt{|g|} g^{\mu\nu} \partial_\nu,
\ea
are given in terms of Hankel functions \cite{Bunch:1978yq}:
\ba\label{solinvlap}
\phi^i_{\mu,\lambda_k}=e^{i\lambda_kx^k}e^{\frac{1-D}{2}t}
H^{(i)}_{\sqrt{\left(\frac{D-1}{2}\right)^2-\mu^2}}(\lambda e^{-t}),
\ea 
where $g$ is defined by (\ref{metric}), $i=1,2$ and $\lambda=\sqrt{\delta^{kl} \lambda_k \lambda_l}$, while $\mu^2$ is the eigenvalue of $\Box_g$.

\section{Representation theory and harmonic analysis on the group $\mathbb{R}^D_\kappa$}
We define the fuzzy $\kappa$-Minkowski space as an algebra ${\cal A}$ of Hilbert-Schmidt operators 
acting on the suitable Hilbert space supplemented with the suitable derivatives such that there exists 
a limit in which ${\cal A}$ reproduces algebra of functions  
on the principal homogeneous space of the group $\mathbb{R}^D_\kappa$ \cite{Grosse:1993uq}.  
Assuming that a regular representation, 
defined as a representation on the Hilbert space of functions over principal homogeneous space,
is decomposable into irreducible modules, 
we  are  interested in the unitary irreducible representations of the group $\mathbb{R}^D_\kappa$.
These can be built using the method of induced representations and utilizing  the aforementioned 
fiber bundle structure of the  Lie group. 

Having at hand a representation $\rho(h),\;h\in H$ of the subgroup $H$ one considers the space $G\times {\cal H}$ and nonintersecting orbits  $(g,\psi) 
\rightarrow  (gh,\rho(h^{-1})\psi),\;h\in H$, with ${\cal H}$ being carrier of $\rho$. With respect to the defined action of $H$,
the space $G\times {\cal H}$ splits into equivalence classes, i.e. the transitivity classes that contain points in the same orbit.   
As a result, the space of transitivity classes can be identified with an associated vector bundle $G/H \times {\cal H}$.
Moreover, the equivalence of points in $G\times{\cal H}$ implies that any section of $G/H\times {\cal H}$ defines 
a section of $G\times {\cal H}$ such that $\psi(g)=\rho(h)\psi(gh),\;\forall h\in H$, that is 
a function $\psi(g)$ on $G$ with values in ${\cal H}$ equivariant with respect to the right action of $H$. 
Then, the action of the group $G$ on $\psi(g)$ induced by left translations 
gives the Mackey's representation of the group $G$ induced by the representation $\rho$ of the subgroup $H$.

The group $\mathbb{R}^D_\kappa$ is a semidirect product of a one-parameter subgroup and 
abelian normal subgroup, and since unitary irreducible representations of both subgroups are one-dimensional,
the relevant induced representations are defined on the sections of appropriate line bundles 
and can be viewed as a representation on the space 
of complex  functions on the $G/H$.
 
Unitary representations $\rho^\Lambda_*,\;\Lambda\in\mathbb{R}^{D-1}$,
induced by one-dimensional unitary irreducible representations of the maximal abelian normal subgroup
\ba\label{odr}
\rho^\Lambda(e^{iy^k\hat{x}_k})|\Lambda\rangle=e^{-i\lambda_ky^k}|\Lambda\rangle,
\ea 
in analogy to two-dimensional case discussed in details in \cite{gn,vil,Agostini:2005mf,Durhuus:2011ci},
are realized on the Hilbert space $L^2(\mathbb{R})$ of square integrable functions on the real line
\ba\label{JSmap}
\rho^\Lambda_*(\hat{x}_0)=-i\partial_Q,\;\rho^{\Lambda}_*(\hat{x}_k)= \lambda_ke^{Q}.
\ea 
By the same reasoning as in \cite{vil}, one concludes that for $\Lambda\in\mathbb{R}^{D-1}/\{0\}$
classes of inequivalent unitary irreducible representations are labeled by points
on the unit sphere $S^{D-2}$ in accordance with \cite{Dabrowski:2010yk}. Specifically, sphere  
$S^0$ is defined as a set with two points in this context. For $\Lambda=0$, the normal subgroup 
is mapped trivially to the unit and therefore induced representation can be 
viewed as a regular representation of the one-dimensional subgroup generated by $\hat{x}_0$.
Since regular representation of any one-parameter Lie group is not irreducible, it follows that 
induced representation for $\Lambda=0$ is not irreducible as well.  
In order to verify these results one should note that the kernel of the homomorphism (\ref{odr}) is isomorphic 
to $\mathbb{R}^{D-2}$. More precisely, it is orthogonal complement in $\mathbb{R}^{D-1}$ of the one-dimensional
space generated by vector $\Lambda$ and therefore the problem of construction of the irreducible representations of $\mathbb{R}^D_\kappa$
reduces to an equivalent problem for $\mathbb{R}^2_\kappa$. Representations characterized by  collinear vectors $\Lambda$ are 
equivalent \cite{vil}. 

Furthermore, following \cite{vil}, it can be shown that unitary representation $l^\omega_*$ induced by the
one-dimensional unitary irreducible representation $l^\omega$ of the one-dimensional subgroup generated by $\hat{x}_0$
\ba\label{irrepx0}
l^\omega(e^{it\hat{x}_0}) |\omega\rangle=e^{-it\omega}|\omega\rangle
\ea 
contains all inequivalent irreducible unitary representations $\rho^\Lambda_*$:
\ba\label{indtreb}
l^\omega_*=\int^{\Lambda \in S^{D-2}}\rho^{\Lambda}_*,
\ea
where trivial representation is omitted from the sum. For $D=2$, integral in (\ref{irrepx0}) has to be replaced by sum over two inequivalent
unitary irreducible representations. 
Denoting the carrier of the one-dimensional representation $l^\omega$ as ${\cal H}_\omega$, the described identification of points of the space 
$G\times {\cal H}_\omega$ that belong to the same orbit under the action of the one-parameter group generated by $\hat{x}_0$ selects only 
Mackey's functions with property
\ba 
\psi(e^{it\hat{x}_0} e^{ix^i\hat{x}_i})=e^{-it \omega} \psi(e^{ie^{t} x^i\hat{x}_i}).
\ea 
Consequently, the representation induced by the left translations as the pull-back $\phi_{\tilde{g}*} \psi$ of the function $\psi$
\ba 
(\phi_{\tilde{g}*} \psi)(g)=\psi({\tilde{g}}^{-1}g),\;\forall \tilde{g} \in G, 
\ea
on the restricted functions $\psi(e^{x^i\hat{x}_i})$ is given by
\ba\label{redone}
&& l^\omega_*(e^{it\hat{x}_0} e^{i\lambda^k\hat{x}_k})\psi(e^{ix^i\hat{x}_i})=
e^{i\omega t}\psi(e^{ie^{-t}(x^i-\lambda^i)\hat{x}_i}),\nn\\
&& l^\omega_*(\hat{x}_0)=\omega+ix^k\partial_k,\;l^\omega_*(\hat{x}_i)=i\partial_i.
\ea 
In order to verify the decomposition (\ref{indtreb}), one introduces the Fourier transformation:
\ba 
{\cal F}\psi  (k_i)=\int d^{(D-1)} x^i e^{ik_lx^l} \psi(x^i),
\ea
and finds the action of the group on the dual space that follows from (\ref{redone})
\ba 
{\cal F}l^\omega_*(e^{it\hat{x}_0} e^{i\lambda^k\hat{x}_k})\psi(e^{ix^i\hat{x}_i})&=&
\int d^{(D-1)} x^i e^{ik_lx^l} e^{i\omega t}\psi(e^{ie^{-t}(x^i-\lambda^i)\hat{x}_i})=\nn\\
&=& e^{i\omega t} e^{ik_l\lambda^l}e^{t(D-1)}{\cal F}\psi (e^tk_i). 
\ea
After redefinition 
\ba 
{\cal F}\psi (k_i)=\prod_{l=1}^{D-1} k_l^{-i\omega-D+1}\phi(k_i),
\ea 
the corresponding action of the group on the functions $\phi$ is given by
\ba\label{newac}  
e^{it\hat{x}_0} e^{ix^k\hat{x}_k}\triangleright \phi(k_i)=e^{ik_lx^l}\phi(e^tk_i)=e^{i|k|\left(\frac{k_lx^l}{|k|}\right)}\phi(e^tk_i),\;
|k|=\sqrt{\sum_{l=1}^{D-1}k_l^2}.
\ea
Hence, we obtained representation on the space of functions on $\mathbb{R}^{D-1}$ given by 
\ba\label{defl}
l_{\cal F*}^0(\hat{x}_0)=-ik_i\tilde{\partial}^i,\;l_{\cal F*}^0(\hat{x}_i)=k_i,\;\tilde{\partial^i}=\frac{\partial}{\partial k_i},
\ea
and denoted by $l_{\cal F*}^0$ to indicate that it is related to $l_*^0$ by Fourier transform.
Comparison of (\ref{newac}) with the action in the irreducible representation (\ref{JSmap})
\ba 
\rho^\Lambda_*(e^{it\hat{x}_0} e^{ix^k\hat{x}_k})\psi(x)=e^{i\lambda_lx^l x}\psi(e^t x),\;x=e^Q>0,
\ea 
shows that to each ray, i.e. half-line diffeomorphic to $\mathbb{R}_+$ 
defined by $(D-1)$-tuple $(k_i,\;|k| \neq 0)$, a unique unitary irreducible representation $(\rho^\Lambda_*,\Lambda \in S^{D-2})$ 
with $\lambda_i=k_i/|k|$  is assigned. Point $|k|=0$ corresponds to a trivial representation.
 
Finally, for $H$ being the trivial subgroup that contains only the unit element, the prescribed procedure results in the left regular representation $L_*$
which can be decomposed as
\ba\label{decomreg} 
L_*=\int_{\oplus}^{\omega\in \mathbb{R}} \;l^\omega_*.
\ea 
As has already been explained at the beginning of this section, regular representation and its structure is of the main interest for our purpose and  
once again, in order to verify decomposition (\ref{decomreg}), one can follow \cite{vil}.
Having at hand a function on principal homogeneous space, written as $\psi(g)=\psi(e^{x^0},x^i)$, define 
function $\psi(te^{x^0},tx^i),\;t\in \mathbb{R}_+$ and its Mellin transform with respect to the $t$ dependence:
\ba\label{mt}
{\cal M}\psi(te^{x^0},tx^i)\equiv \psi^\sigma (e^{x^0},x^i)=\int_0^\infty dt \psi(te^{x^0},tx^i) t^{-\sigma-1}.
\ea
Those definitions ensure that the action of the group $\mathbb{R}_\kappa^D$ on  (\ref{mt}) derived from the left regular representation 
coincides with the representation (\ref{redone}) with $\sigma=i\omega$. Accordingly, the Mellin transform (\ref{mt}) restricts 
the regular representation to particular components (\ref{redone}), while the inverse of the Mellin transform
\ba
\psi(te^{x^0},tx^i)=\frac{1}{2\pi i} \int_{c-\infty}^{c+\infty} d\sigma \psi^\sigma(e^{x^0},x^i) t^\sigma,
\ea 
for the  particular value $t=1$, explicitly reproduces (\ref{decomreg}). The same result can be inferred by considering the expansion 
of the scalar field on the principal homogeneous space of the group $\mathbb{R}_\kappa^D$ into the eigenfunctions (\ref{solinvlap}).
There, the two roots of the eigenvalue $\mu^2=(\pm\omega)^2$ correspond to the two-fold degeneracy $(i=1,2)$ of Hankel functions, while spacelike plane waves 
correspond to the decomposition (\ref{indtreb}).

\section{Fuzzy $\kappa$-Minkowski space}  
In an attempt to construct a noncommutative variant of homogeneous spaces 
it is convenient, as in  \cite{Grosse:1993uq}, to realize the regular representation $L_*$ by the action of the group on the group algebra, i.e. an algebra with elements of the form
\ba\label{ga}
F[f]=\int \mu_L(g) f(g) g, 
\ea  
where $\mu_L(g)$ is left invariant measure, while $f(g)$ is any finite distribution 
with compact support over smooth functions on the group manifold.
The action of the group on (\ref{ga}) is defined by left multiplication and the 
structure of the algebra is ensured by defining the multiplication of elements (\ref{ga})  
using the multiplication in the group
\ba 
F_1F_2=\int \mu_L(g) \mu_L(g^\prime)f_1(g) f_2(g^\prime) gg^\prime. 
\ea
This multiplication can be realized on the algebra of functions/distributions by replacing the
usual pointwise product with the convolution
\ba\label{convol} 
(f_1 \star f_2)(g)=\int \mu_L(g^\prime) 
f_1({g^\prime}) f_2({g^\prime}^{-1}g). 
\ea
Supplemented with the involution defined by modular function $\Delta$
\ba 
f^*(g)=\bar{f}(g^{-1})\Delta(g^{-1}),
\ea 
where bar denotes complex conjugation,
the completion of the group algebra in the $L^1(G)$ norm is 
a Banach $*$-algebra, isomorphic to the $L^1(G)$ as a vector space.

The structure of the $\mathbb{R}^2_\kappa$-group algebra has been studied in Refs.\cite{Agostini:2005mf,Durhuus:2011ci}, where 
it has been shown that, although generically $f(g)$ in (\ref{ga}) is regarded as an element of the $L^1(\mathbb{R}^2_\kappa)$,
the set ${\cal B}\cap L^2(\mathbb{R}_\kappa^2)$ is dense in $L^2(\mathbb{R}_\kappa^2)$, 
where Schwartz space ${\cal B}\subset L^1(\mathbb{R}_\kappa^2)$ is defined as a Fourier dual of the 
Schwartz space ${\cal S}_c$ of the functions with compact support in the time-like variable.
This result, extensible to $\mathbb{R}_\kappa^D$, enabled the authors of \cite{Durhuus:2011ci} to prove 
that an element of the group algebra is of the Hilbert-Schmidt type if and only if the function $f(g)$ is a Fourier transform of some square 
integrable function. Moreover, it enabled them to define a
star product of the functions on $L^2(\mathbb{R}_\kappa^2)$,
as well as a quantization map which resembles Weyl quantization of phase space. However, we are not only interested in the 
convolution/star product of functions, but also in the construction of fuzzy variant of the Laplace-Beltrami operator (\ref{invlap}).
Therefore we need to define fuzzy variants of vector fields and, seemingly, 
the easiest way to achieve this is in the coherent states approach elaborated in \cite{Grosse:1993uq}. 
In the following, we restrict discussion on the subset of the $\mathbb{R}^D_\kappa$-group algebra that contains only elements of the Hilbert-Schmidt
type.

Having at hand a Lie group $G$, coherent states are defined as an overcomplete set of states that belong to 
an unitary irreducible representation $\pi$ of the group \cite{Perelomov}. The set of coherent states can be 
built by the action of the group
\ba 
|\phi_g\rangle= \pi(g)|\phi_0\rangle,
\ea 
on any admissible initial state $|\phi_0\rangle$ from the G\aa rding domain of representation. 
If there is a so-called stability subgroup $H$ of the group $G$ characterized by 
\ba 
\pi(h)|\phi_0\rangle=e^{i\alpha(h)}|\phi_0\rangle,\;\alpha(h) \in\mathbb{C},\;\forall h\in H,
\ea 
then states $|\phi_g\rangle$ and $|\phi_{hg}\rangle$ are equivalent.
Moreover, the representation of the group has an obvious extension to the representation of the group algebra and
the set of coherent states topologically
coincides with  $G/H$. This enables to assign a function on the homogeneous space $G/H$    
to any element of the group algebra (see e.g. \cite{Grosse:1993uq} and references therein) 
\ba 
\tilde{f}(P)={\cal Q}^{-1} (F)=\langle \phi_P| \pi(F)|\phi_P \rangle,\;P\in G/H,
\ea
where the state $|\phi_P\rangle$ is obtained by the action of any representative of the coset $gH$ on the
initial state $|\phi_0\rangle$.
Notation ${\cal Q}^{-1}$ suggests that the map has to be interpreted as the inverse of the quantization map. 
Using the dequantization map ${\cal Q}^{-1}$, one can define a star product of the 
functions on the homogeneous space $G/H$ 
\ba 
{\cal Q}^{-1}(\pi( F_1))\hat{\star}{\cal Q}^{-1}(\pi(F_2))={\cal Q}^{-1}(\pi(F_1F_2)). 
\ea
If the representation $\pi$ is faithful, i.e. if the subgroup $H$ is trivial, 
then the product $\hat{\star}$ is equivalent to the  
convolution (\ref{convol}). Otherwise, the space of functions on $G/H$ obtained by 
the dequantization map supplemented with $\hat{\star}$-product has to be interpreted as the representation of the convolution 
algebra.

 To get better insight in the application of the described formalism to $\mathbb{R}^D_\kappa$ group, 
 first we restrict discussion to $D=2$ and then we extend results to 
 higher dimensions incorporating appropriate modifications.
 Of the particular interest is the embedding of the group algebra of $\mathbb{R}^2_\kappa$ 
 into the group algebra of the Heisenberg group.

Recall that two-dimensional $\kappa$-Minkowski space admits only two inequivalent unitary irreducible representations.
Both of them can be realized  on the unitary irreducible representation of the Heisenberg group 
 by the Jordan-Schwinger map (\ref{JSmap}) \cite{Agostini:2005mf,Pachol:2015qia}. More precisely, 
 the carrier of the Schr{\"o}edinger representation of Heisenberg group can be viewed as a representation space of 
 the representation $l^0_{{\cal F}*}$ described in previous section. We showed that the representation $l^0_{{\cal F}*}$ 
 can be decomposed into the direct sum of the trivial one, assigned to the point $x=0$ on the real line,
 and two inequivalent irreducible representations, defined on the space of functions non-vanishing either only on the strictly positive or on the strictly negative half-line. 
 On the other hand, assuming an irreducible representation of the Heisenberg group, a convenient matrix basis 
 for the Hilbert-Schmidt operators is defined in terms of states 
 generated by a successive action of the creation operators $a^\dagger$ on the 
 "ground" state $|0\rangle$ annihilated by $a$, where 
 $[a^\dagger,a]=-1$ as usual (see for example \cite{Lizzi:2014pwa} and references therein).
 Therefore, due to the aforementioned decomposition, elements of the $\mathbb{R}^2_\kappa$-group algebra, considered as infinite-dimensional matrices
 embedded into the group algebra of the Heisenberg group, split into the 
 two infinite-dimensional blocks. 
 Choosing an initial normalizable state $|\phi_0\rangle$ from the carrier of an non-trivial irreducible component,
 we define the collection of states 
 \ba\label{pscs} 
 |t,y^1\rangle=e^{it\hat{x}_0}e^{iy^1\hat{x}_1}|\phi_0\rangle.
  \ea
 In the case of the Heisenberg group, it is convenient to specify a unique initial state 
 by imposing holomorphicity, but for $\kappa$-Minkowski space such a condition is not suitable as has been 
 discussed in \cite{klauder,Aslaksen:1969dy} where an acceptable choice is suggested. 
 Since the collection of states (\ref{pscs}) is defined  with respect to an irreducible component of the representation $l^0_{{\cal F}*}$,
 it should be clear that states (\ref{pscs}) represent an overcomplete set of states
 with respect to the half of the representation space of  $l^0_{{\cal F}*}$.
 Furthermore, it is well known that Schr{\"o}edinger representation does not admit normalizable eigenstate of any element  
 of the Lie algebra of the Heisenberg group. Similarly, in a unitary irreducible representation of $\mathbb{R}^2_\kappa$ there 
 is no normalizable eigenstate of any non-trivial element of the Lie algebra generators of $\mathbb{R}^2_\kappa$ as well.
 Therefore, the stability group related to the collection of states (\ref{pscs}) is trivial, thus implying that the representation of the group
$\mathbb{R}^2_\kappa$ on the space spanned by the collection of states (\ref{pscs}) is faithful. 
  
 Bearing in mind  results obtained for $D=2$, in higher dimensional case we define coherent states with respect 
 to the representation $l^0_{{\cal F}*}$. Therefore, assuming the representation $l^0_{{\cal F}*}$,
 the representation space is the space of functions on $\mathbb{R}^{D-1}$.
 It can be viewed as a tensor product of $(D-1)$ copies of the representation space of the Schr{\"o}edinger representation of Heisenberg group,
 where each copy admits the representation of $\mathbb{R}^2_\kappa$ group decomposable into two non-equivalent
 unitary irreducible components. In analogy to $\mathbb{R}^2_\kappa$, we select one of these non-equivalent representations for 
 each component in the tensor product. For the sake of definition, in each component we select space of functions
 non-vanishing for strictly positive half-line. Thus we obtain the representation on the space of functions non-vanishing 
 only on the subset $x_i>0,\;i=1,\cdots ,D-1$ of $\mathbb{R}^{D-1}$. Furthermore, assuming a suitable initial state $|\phi_0\rangle$,
 we define a collection of states by the action of the group $\mathbb{R}^D_\kappa$
 \ba\label{cohdd}
 |t,y^1,\cdots,y^{D-1}\rangle=e^{it\hat{x}_0} e^{iy^i\hat{x}_i}|\phi_0\rangle.
 \ea 
 In order to show that the collection of states (\ref{cohdd})
 represents an overcomplete set of states with respect to the selected representation space,
 we define the operator  
 \ba 
 B=\int \mu_L(t,y^1,\cdots,y^{D-1})  |t,y^1,\cdots,y^{D-1}\rangle\langle t,y^1,\cdots,y^{D-1}|,
 \ea 
 which commutes with all operators $l_{\cal F*}^0(g)$.  
 By virtue of (\ref{newac}) and (\ref{defl}), $l^0_{{\cal F}*}$ is decomposable 
 into irreducible components   
 \ba\label{deco222}
 l^0_{{\cal F}*}=\int^{\Lambda \in S^{D-2}} \rho^\Lambda_*,
  \ea
  and it follows that $B$ is non-vanishing only on the portion $(1/2)^{D-1}$ of the $S^{D-2}$ sphere in (\ref{deco222}).
  Moreover, if we impose the condition $\phi_0(x_i)=\phi_0(|x|)$ on the function that corresponds to the initial state $|\phi_0\rangle$,
  then the operator $B$ is proportional to unit on the portion of domain on which it does not vanish, thus confirming 
  that the collection of states (\ref{cohdd}) is an overcomplete set of states.
  As a final remark on the properties of the representation constructed by embedding of $\mathbb{R}^D_\kappa$-group algebra 
  into the group algebra of Heisenberg group, we point out that the representation on the Hilbert space spanned by the 
  collection of states (\ref{cohdd}) is faithful.
  
  Finally, with the overcomplete collection of states which span the faithful representation at hand, we define an isomorphism 
  from the group algebra to the space of functions on $\mathbb{R}^D_\kappa$
  \ba\label{isoga}
 {\cal Q}^{-1}(F)=\langle \phi_0|e^{-iy^i\hat{x}_i}e^{-it\hat{x}_0}F
 e^{it\hat{x}_0} e^{iy^i\hat{x}_i}
 |\phi_0 \rangle.
 \ea 
 Supplemented with the product 
 \ba 
 {\cal Q}^{-1}(F_1)\hat{\star}{\cal Q}^{-1}(F_2)={\cal Q}^{-1}(F_1F_2),
 \ea 
 the space of functions on $\mathbb{R}^D_\kappa$ has a structure of an algebra isomorphic to the 
 convolution algebra. Isomorphism depends on the choice of the initial state, as is evident from 
 \ba 
 {\cal Q}^{-1}(F)=\int \mu_L(g) f(g) \langle \phi_0|e^{-iy^i\hat{x}_i}e^{-it\hat{x}_0} g
  e^{it\hat{x}_0} e^{iy^i\hat{x}_i}
 |\phi_0 \rangle.
 \ea 
 Furthermore, we define derivatives as  commutators with the Lie algebra generators, thus ensuring Leibnitz rule. Taking into account the explicit form of the 
 Laplace-Beltrami operator (\ref{invlap}) 
 \ba\label{classdesitt}
(\partial_t^2+(D-1)\partial_t+\mu^2-e^{2t}\delta^{ij}\partial_i\partial_j)f(t,y^i)=0
\ea
 and the identities
  \ba\label{comder} 
  && i\partial_t {\cal Q}^{-1}(F)(t,x^1,\cdots,x^{D-1})={\cal Q}^{-1}([\hat{x}_0,F])(t,x^1,\cdots,x^{D-1}),\;\nn\\
  && ie^{-t}\partial_{j} {\cal Q}^{-1}(F)(t,x^1,\cdots,x^{D-1})={\cal Q}^{-1}([\hat{x}_0,F])(t,x^1,\cdots,x^{D-1}),
 \ea
 easily derived using (\ref{isoga}) and (\ref{liealg}), we define fuzzy the Laplace-Beltrami 
 operator as 
 \ba\label{fuzzlap}
\hat{\Box}_g \equiv -[\hat{x}_0,[\hat{x}_0,\;]]-i(D-1)[\hat{x}_0,\;]+\mu^2+\delta^{ij}[\hat{x}_i,[\hat{x}_j,\;]]=0. 
\ea
 It follows that the map ${\cal Q}^{-1}$ assigns to an eigenvector of fuzzy Laplacian $\hat{\Box}_g$ an eigenfunction of the
classical Laplace-Beltrami operator $\Box_g$.
 
\section{Conclusion}
We considered the Lie group $\mathbb{R}^D_\kappa$ generated by the Lie algebra of $\kappa$-Minkowski space.
Imposing the invariance of the metric under the pull-back of diffeomorphisms induced by right translations in the group $\mathbb{R}^D_\kappa$, we found that a
unique right invariant metric is associated with $\mathbb{R}^D_\kappa$.
This metric coincides with the metric of de Sitter space-time in planar coordinates.
Consequently, principal homogeneous space of the group $\mathbb{R}^D_\kappa$ endowed with right invariant metric coincides with the half of de Sitter space-time.
The automorphism $g\to g^{-1}$ of the group $\mathbb{R}^D_\kappa$ implies the same conclusion for the left invariant metric. 

Furthermore, we presented an analysis of the structure of unitary representations of the Lie group 
$\mathbb{R}^D_\kappa$ which are relevant for the formulation of the non-commutative $\kappa$-Minkowski space 
in matrix basis. From the practical point of view, a suitable matrix basis can be selected using 
an embedding of the $\mathbb{R}^D_\kappa$-group algebra into the group algebra of 
$(2D-1)$-dimensional Heisenberg group while faithfully represented on the space $L^2(\mathbb{R}^{D-1})$, 
as suggested in \cite{Pachol:2015qia}.

An insight in the particular properties of the embedding resulted in the construction of the specific collection of states
and an associated operator $B$ dependent on the choice of an initial state $|\phi_0\rangle$. 
With a suitable conditions imposed on the initial state, the operator $B$ can be interpreted as a projector
from the space $L^2(\mathbb{R}^{D-1})$ to the subspace $L^2(\mathbb{R}_+^{D-1})$ for which the constructed collection of states 
realizes an overcomplete set of states. Space $L^2(\mathbb{R}_+^{D-1})$ carries a reducible representation of the group $\mathbb{R}^{D-1}_\kappa$
and therefore the constructed collection of states can be considered as a generalization of the coherent states defined in \cite{Perelomov}.

Using the constructed overcomplete set of states 
and the fact that aforementioned reducible representation of the group $\mathbb{R}^{D-1}_\kappa$ on $L^2(\mathbb{R}_+^{D-1})$
is faithful, we defined the dequantization map, that is an isomorphism from the space of Hilbert-Schmidt
operators acting on $L^2(\mathbb{R}_+^{D-1})$ onto to space of functions $L^2(\mathbb{R}^{D-1})$.
Although the dequantization map itself depends on the choice of initial state  $|\phi_0\rangle$, a star product 
naturally induced by dequantization map is equivalent to convolution. 
 
Finally, we defined the fuzzy variant of the Laplace-Beltrami operator and we showed that 
dequantization map relates the fuzzy eigenfunctions with the eigenfunctions that correspond to the same eigenvalue of the classical 
Laplace-Beltrami operator on de Sitter space-time. In order to define the dequantization map more explicitly 
it remains to solve the fuzzy eigenvalue equations or to choose aforementioned initial state in some particular way which 
will be discussed elsewhere together with the analysis of the semi-classical limit 
and an action for the field theory on the fuzzy $\kappa$-Minkowski space.  

Comparing the presented approach with the similar approaches to the 
construction of non-commutative spheres and their pseudo-Riemaniann counterparts
\cite{Madore:1991bw, Grosse:1996mz, Ho:2000fy, Guralnik:2000pb, Ramgoolam:2001zx, 
SheikhJabbari:2004ik, Abe:2004sa, SheikhJabbari:2005mf,Hasebe:2010vp,Hasebe:2012mz,
Jurman:2013ota, Sperling:2017gmy, Sperling:2017dts, Buric:2017yes},
one should note that the presented approach determines a map only onto the half of de Sitter space.

{\bf Note added.} With respect to this comparison, the presented approach utilizes the principle bundle structure of solvable group generated by 
$\kappa$-Minkowski Lie algebra to define an embedding of fuzzy half of de Sitter space into the higher dimensional phase space.
In this context, by embedding of a fuzzy space into the higher dimensional phase space we mean realization of the non-commutative algebra which 
defines the fuzzy space as a subalgebra of the Heisenberg algebra. Such a realization for fuzzy principal homogeneous spaces of a general Lie group 
and the corresponding star product has been discussed in \cite{Chryssomalakos:2007jr}. Similar construction for homogeneous spaces   
$G/H$, for a general Lie group $G$ and a continuous subgroup $H$, with a principal bundle structure 
encoded in the short sequence $H\to G\to G/H$, is still missing. The exceptions are   
homogeneous spaces which descend from classical groups, e.g. complex projective space, Euclidean spheres $SO(n)\to SO(n+1)\to S^n$ or de 
Sitter spaces $SO(n-1,1)\to SO(n,1)\to dS^n$. For Euclidean spheres $SO(n+1)/SO(n)$ of generic dimension, for example, such an embedding is given in 
\cite{SheikhJabbari:2004ik,SheikhJabbari:2005mf,SheikhJabbari:2006bj} with a special attention devoted to    
spheres related to Hopf fibrations. In those references, embedding is realized using Clliford algebra and constraints providing 
a projection from Hilbert space of Heisenberg group to fuzzy $S^n_F$ are discussed. In the light of this, in an attempt to 
extend the construction presented in this paper to full de Sitter space, a better understanding of the relation with the construction of fuzzy 
spheres is desirable, especially a relation between fuzzy two-dimensional $\kappa$-Minkowski space with 
the coherent state approach to fuzzy sphere  \cite{Grosse:1996mz,Hammou:2001cc}.

Finally, apart from providing an insight in the structure of the non-commutative 
$\kappa$-Minkowski space from the matrix geometry point of view  
the presented construction might be implemented into the group theory approach 
to quantum field theory on de Sitter space-time \cite{Joung:2006gj,Joung:2007je}.

\section*{Acknowledgments}
The author is grateful to A. Chatzistavrakidis, T. Juri\'c and  Z. \v Skoda for the encouraging and 
valuable discussions.  
The work was supported by the Croatian Science Foundation under the project IP-2014-
09-3258, as well as by the H2020 Twinning project No. 692194,"RBI-T-WINNING".
For the additional note in relation to fuzzy spheres and for bringing attention to some of the references added,
the author acknowledges M.M. Sheikh-Jabbari.

\end{document}